\documentclass[aps,superscriptaddress,nofootinbib,eqsecnum,prd,notitlepage,twocolumn]{revtex4-1} 

\pdfoutput=1

\usepackage{amsfonts}
\usepackage{amsmath}
\usepackage{amssymb}
\usepackage{graphicx,color}
\usepackage{float}
\usepackage{hyperref}
\usepackage{subfigure}
\usepackage{dcolumn}
\usepackage{soul}
\usepackage{ulem}
\usepackage{verbatim}

\begin{document}

\title{Runaway potentials in warm inflation satisfying the swampland
  conjectures}

\author{Suratna Das}
\email{suratna@iitk.ac.in}
\affiliation{Department of Physics, Indian Institute of Technology,
   Kanpur 208016, India}

\author{Rudnei O. Ramos} \email{rudnei@uerj.br}
\affiliation{Departamento de Fisica Teorica, Universidade do Estado do
  Rio de Janeiro, 20550-013 Rio de Janeiro, RJ, Brazil }

\begin{abstract}

The runaway potentials, which do not possess any critical points, are
viable potentials which befit the recently proposed de Sitter
swampland conjecture very well. In this work, we embed such potentials
in the warm inflation scenario motivated by quantum field theory
models generating a dissipation coefficient with a dependence cubic in 
the temperature. It is demonstrated that such models are able to remain in
tune with the present observations and they can also satisfy all the three
Swampland conjectures, namely the Swampland Distance conjecture, the
de Sitter conjecture and the Transplackian Censorship Conjecture,
simultaneously. These features make such models viable from the point of view
of effective field theory models in quantum gravity and string theory, 
away from the Swampland. 

\end{abstract}

\maketitle

\section{Introduction}

The fact that achieving a stable or meta-stable de Sitter vacuum in M
or String theory has proven to be  a tasking job for decades (see,
e.g., Ref.~\cite{Danielsson:2018ztv} for a recent
review)\footnote{
  Note that despite
  existing proposals for constructing de Sitter states in string theory,
  like, e.g., the KKLT scenario~\cite{Kachru:2003aw}, there are still concerns about that
  (see, e.g., Ref.~\cite{Sethi:2017phn}). {}For a recent discussion about the difficulties
  in constructing de Sitter states in string theory, see, for instance,
  Ref.~\cite{Dine:2020vmr}.}, has led to the
recently proposed de Sitter Swampland criterion~\cite{Obied:2018sgi,
  Garg:2018reu, Ooguri:2018wrx} which is conjectured to constrain
constructions of the de Sitter vacuum within String landscapes.  This
conjecture has created a lot of discussion in the recent literature
and for good reasons, since the early and the late time cosmologies
are in need of phases where the universe evolves like close to a de Sitter
state~\cite{Agrawal:2018own, Agrawal:2018rcg}. Another decade old
Swampland criterion, known as the Swampland distance
conjecture~\cite{Ooguri:2006in} and which was formulated to restrict
tower of masses to appear in a low-energy effective field theory, has
cosmological implications as well and where the cosmological dynamics involves
evolution of scalar fields, like inflation. The latest addition to
this list of Swampland criteria is the Transplanckian Censorship
Conjecture (TCC)~\cite{Bedroya:2019snp, Bedroya:2019tba}, which has
been devised to refrain any sub-Planckian primordial modes from
leaving the causal horizon during the inflationary phase and which would seed
the structures in our universe at later stages. 

In short, all the three above mentioned Swampland criteria restrict
the dynamics of inflation in one way or another. The de Sitter
conjecture, which puts bounds on the slope of the scalar potentials in
an effective field theory~\cite{Garg:2018reu, Ooguri:2018wrx},
\begin{eqnarray}
|\nabla V|\geq \frac{c}{M_{\rm Pl}}V, \quad {\rm or}\quad {\rm
  min}(\nabla_i\nabla_j V)\leq - \frac{c'}{M_{\rm Pl}^2}V,
\end{eqnarray}
where $c$ and $c'$ are both constants of order unity, essentially
restricts the slow-roll parameters to become smaller than
unity~\cite{Garg:2018reu},
\begin{eqnarray}
\epsilon_V\equiv \frac{M_{\rm
    Pl}^2}{2}\left(\frac{V,_\phi}{V}\right)^2\geq \frac{c^2}{2},\quad
        {\rm or}\quad \eta_V\equiv M_{\rm
          Pl}^2\frac{V,_{\phi\phi}}{V}\leq-c',\nonumber\\
\end{eqnarray}
with $M_{\rm Pl} \equiv 1/\sqrt{8 \pi G} \simeq 2.4 \times 10^{18}$GeV
is the reduced Planck mass.  However, it is essential to have the
slow-roll parameters much smaller than unity for inflation to take
place in a canonical cold inflationary paradigm\footnote{
  An analysis presented in Ref.~\cite{Kinney:2018nny} 
has shown how the above conjectures put strong constraints in cold inflation models
as far the cosmic microwave background data is concerned. }.
The Swampland
distance conjecture, which restricts the excursion of a scalar field in
an effective field theory~\cite{Ooguri:2006in},
\begin{eqnarray}
\frac{\Delta\phi}{M_{\rm Pl}}<d,
\end{eqnarray}
where $d$ is another constant of order unity, essentially favors the
small field models of inflation over the large field ones. On the
other hand, the TCC, which is a bound on the length scales leaving the
causal horizon~\cite{Bedroya:2019tba},
\begin{eqnarray}
\left(\frac{a_f}{a_i}\right)\ell_{\rm Pl}<\frac{1}{H_f},
\end{eqnarray}
where $a_i$ and $a_f$ are, respectively, the scale factors at the beginning and at
the end of the evolution, $H_f$ is the Hubble parameter at the end of
that evolution and $\ell_{\rm Pl}$ is any length scale of order Planck
scale, yields a bound on the scale of inflation\footnote{A modified
  version of the TCC has recently been suggested in
  Ref.~\cite{Aalsma:2020aib} and which proposes that $N$ is bounded only by
  the logarithm of the de Sitter entropy, i.e.,  $N < 2 \ln (M_{\rm Pl}
  / H_f)$. This allows for larger values of $H$ during inflation, $H_f
  / M_{\rm Pl} < 10^{-12}$, which substantially alleviates the bound
  in Eq.~(\ref{TCC-bound}) by some four orders of magnitude (see also
  Refs.~\cite{Mizuno:2019bxy, Kamali:2019gzr, Berera:2020dvn} for other discussions on how
  to relax the TCC bound).} ,
\begin{eqnarray}
V^{1/4} < 6\times 10^8 \,{\rm GeV}\sim 3\times 10^{-10} M_{\rm Pl}.
\label{TCC-bound}
\end{eqnarray}
If inflation takes place at such low-energy scales
then the observed scalar power amplitude can only be 
obtained in cold inflation if $\epsilon_V\sim10^{-31}$, which yields a too 
small tensor-to-scalar ratio ($r=16\epsilon\sim 10^{-30}$) to be detected by
any of the near-future observation \cite{Bedroya:2019tba}. 
Moreover, in such a scenario, one requires $\eta_V\sim\mathcal{O}(10^{-2})$ 
so that the observed scalar spectral tilt ($n_s-1=2\eta_V-6\epsilon_V\sim 2\eta_V$)
can be explained~\cite{Das:2019hto}. It is rather impossible to construct 
a potential which yields $\epsilon_V\sim10^{-31}$ and $\eta_V\sim10^{-2}$ 
so that cold inflation can be made in tune with the 
Transplanckian Censorship Conjecture.  
Therefore, as it is essential to fulfil all the above three Swampland criteria
in order to realize an inflationary phase in any effective field
theory consistent with the String Landscape, it turns out to be a
challenging task to realize canonical single-field slow-roll
inflationary dynamics in a String vacuum~\cite{Kehagias:2018uem,
  Das:2018hqy, Bedroya:2019tba, Das:2019hto}. 

It was pointed out in Refs.~\cite{Das:2018hqy, Motaharfar:2018zyb,
  Das:2018rpg, Rasheed:2020syk} that warm inflation
(WI)~\cite{Berera:1995ie}, a variant inflationary paradigm to the
standard cold inflation scenario, can accommodate the de Sitter
conjecture quite easily due to its very construction. In particular,
the de Sitter conjecture was explicitly analyzed in the WI context in
Ref.~\cite{Brandenberger:2020oav}, where it was demonstrate that this
conjecture remains robust in WI.  In WI, the inflaton field can
continuously transfer its energy to a radiation bath during inflation,
inducing an extra frictional term in the inflaton dynamics and
resulting the field to roll slower than in the cold paradigm (for
reviews on WI, see, e.g.,
Refs.~\cite{Berera:2008ar,BasteroGil:2009ec}).  In that case, WI takes
place when the slow-roll parameter $\epsilon_V$ and $\eta_V$ are
smaller than $1+Q$, with $Q$ being the ratio of the frictional terms
in the inflaton dynamics due to the dissipation, denoted by
$\Upsilon$, and the expansion of the universe, $Q \equiv
\Upsilon/(3H)$, which can be much greater than unity. Therefore, WI
can easily take place with steeper potentials ($\epsilon_V>1$,
$\eta_V>1$) (which the cold paradigm fails to attain) and, thus, WI
can satisfy the de Sitter conjecture with ease. However, it was shown
in Refs.~\cite{Das:2019hto, Berera:2019zdd} that in order to maintain
the TCC, the scale of WI has to be as low as in the cold paradigm (as
given in Eq.~(\ref{TCC-bound})). 

Since then, at least two attempts have been made to construct viable
WI models with steep potentials which can be realized in String
Landscapes by satisfying all the three Swampland conjectures mentioned
above. The first one~\cite{Kamali:2019xnt} was constructed in the
Randall-Sundrum braneworld scenario where the dissipative coefficient
was taken to be depended on both the inflaton field and the
temperature of the radiation bath existing during inflation, and the
steep potential was considered to be of the exponential form,
\begin{eqnarray}
V(\phi)=V_0e^{-\alpha\phi/M_{\rm Pl}}.
\label{expV}
\end{eqnarray}
Such a steep potential in cold inflation leads to power-law type
inflation~\cite{Liddle:1988tb} where inflation does not exit
gracefully in standard general relativity. However, as has been recently shown in
Ref.~\cite{Das:2020lut}, such a potential can gracefully exit in WI if
the dissipative coefficient has a dependence on the temperature of the
radiation bath like $\Upsilon \propto T^p$, with power  $p> 2$. This
was for instance the case studied in the model of
Ref.~\cite{Kamali:2019xnt}, where the dissipation coefficient had a
$T^3$ dependence on the temperature\footnote{Note that, as also shown
  in Ref.~\cite{Das:2020lut}, an additional dependence on the inflaton
  field amplitude does not affect this conclusion.}.  In another
recent study~\cite{Goswami:2019ehb}, the potential~(\ref{expV}) was
also studied in the context of the WI model proposed by the authors
of Ref.~\cite{Berghaus:2019whh}. In Ref.~\cite{Berghaus:2019whh}, a WI
model, namely the Minimal Warm Inflation (MWI) model,  was constructed
where the inflaton was an axion-like field coupled to gauge bosons in
the usual way and whose derived dissipation coefficient turned out to
be of the form $\Upsilon \propto T^3$. In the study done in
Ref.~\cite{Goswami:2019ehb}, which is the second study where WI with
steep potentials has been put to the test against the Swampland
conjectures, it was embed the runaway exponential potential (\ref{expV}) in
the MWI model. However, the authors of that work have shown that such
a combination yields too much red-tilt in the scalar spectrum to be in
accordance with the observations\footnote{A steep runaway potential of the
  type $V(\phi)=V_0[1+\exp(-\alpha\phi/M_{\rm Pl})]$ was also studied
  in Ref.~\cite{Goswami:2019ehb} when embedding it in the MWI
  model. However, it was shown that although such combination can
  satisfy all the three Swampland conjectures while being in
  accordance with observations, inflation fails to the exit gracefully
  within the parameter range studied.} 

The aim of the present paper is to study a generalized form of the runaway
potential given by~\cite{Geng:2015fla, Geng:2017mic, Ahmad:2017itq,
  Lima:2019yyv}
\begin{eqnarray}
V(\phi)=V_0e^{-\alpha(\phi/M_{\rm Pl})^n}.
\label{pot}
\end{eqnarray}
with $n> 1$. We note that with $n>1$ there is no graceful exit problem even for
the case of cold inflation.  {}Furthermore, as shown in
Ref.~\cite{Goswami:2019ehb} that the $n=1$ case produces a way too
red-tilted spectrum in WI with $p>2$ (in the standard general relativity context),  
this will compel us to go beyond $n=1$ and study the cases with $n>1$.    
One study of the generalized
potentials of the form of Eq.~(\ref{pot}) has been performed recently
in Ref.~\cite{Lima:2019yyv} in the WI context and as a quintessential
inflation model. However, in that reference only the weak dissipative
regime of WI, $Q \ll 1$, has been analyzed.  Motivated by the previous
studies indicating that WI in the strong dissipative regime can be
consistent with the swampland criteria, in the present work we
reconsider this type of models in this regime of WI.  It is worth
recalling here that constructing WI models in the strong dissipative
regime has been historically a
challenge~\cite{Bastero-Gil:2019gao}. Here we will show that the model
(\ref{pot}) can support strong dissipation with the well motivated
type of dissipation coefficients behaving like $\Upsilon \propto T^3$,
but also lead to a dynamics that is consistent both from the
observational as well as from the effective field theory (as defined by the
swampland program) point of views.

This paper is organized as follows. In Sec.~\ref{background}, we
briefly review the generics of the WI dynamics for completeness.  In
Sec.~\ref{sec3} we present some useful analytical studies to determine
the field ranges which are suitable for our analysis and used in the
subsequent section. In Sec.~\ref{numerical} we then perform a full
numerical study covering the appropriate parameter ranges leading to a
consistent inflationary dynamics in the strong dissipative regime.
Our conclusions are presented in the Sec.~\ref{conclusions}.
Two appendices are also included to describe some of the technical
details.

\section{Brief review of Warm inflation}
\label{background}

{}First, let us briefly review the background dynamics of a generic WI
paradigm. In WI, the inflaton dissipates its energy to a constant
radiation bath throughout inflation. Thus, the background dynamics
involve the evolutions equations for the inflaton field $\phi(t)$, for
the radiation energy density $\rho_R(t)$ (or, equivalently, for the
temperature $T(t)$ of the thermal bath as $\rho_R\propto T^4$) and the
{}Friedmann equation, which accounts for the evolution of the scale
factor,
\begin{eqnarray}
&&\ddot\phi+3H\dot\phi+V,_{\phi}=-\Upsilon(\phi,T)\dot\phi,\\ &&
  \dot\rho_R+4H\rho_R=\Upsilon(\phi,T)\dot\phi^2,\\ &&
  H^2=\frac{1}{3M_{\rm
      Pl}^2}\left[\frac{\dot\phi^2}{2}+V(\phi)+\rho_R\right].
\label{full-eqns}
\end{eqnarray}
Here, $\Upsilon$ is the rate of dissipation at which the inflaton
decays to the radiation bath. In general, $\Upsilon$ can be a function
of both $\phi$ and $T$. Some details of the derivation of these
dissipation coefficients in the context of WI have been given in the
Appendix~\ref{appA}. The dimensionless ratio of the two frictional
terms in the inflaton equation of motion, the one due to dissipation
and the other due to the expansion of the universe, is defined as
\begin{eqnarray}
Q\equiv \frac{\Upsilon(\phi, T)}{3H},
\end{eqnarray}
which broadly classifies WI models into two classes: WI taking place
in the weak dissipative regime, where $Q\ll 1$,  and WI taking place
in the strong dissipative regime, when $Q\gg 1$. The slow-roll
parameters in WI are modified with respect to the ones in the cold
inflation scenario to
\begin{eqnarray}
\epsilon_{WI} &=& \frac{\epsilon_V}{1+Q} ,
\label{eps}
\\ \eta_{WI} &=& \frac{\eta_V}{1+Q},
\label{eta}
\end{eqnarray} 
and WI ends when $\epsilon_V\sim 1+Q$. The fact that during inflation
the energy density would be dominated by the potential energy density
of the  inflaton field, and the radiation bath produced would be of
(approximately) constant energy density helps us to reduce the above
dynamical equations to the approximate ones,
\begin{eqnarray}
&&3H(1+Q)\dot\phi\approx V,_{\phi}, \\ && \rho_R\approx
  \frac{3Q}{4}\dot\phi^2\\ &&H^2\approx \frac{V}{3M_{\rm Pl}^2},
\label{approx-eqns}
\end{eqnarray}
where we must note that standard slow-roll approximations, like
$\epsilon_V\ll 1$ or $\eta_V\ll 1$, have not been employed in getting
the above approximated results. Hence, these approximations hold true
even in cases of steep potentials for which $\epsilon_V>1$ and/or
$\eta_V>1$ and, yet, an inflationary regime can still be supported
provided that $Q$ is large enough.

Let us now briefly discuss the perturbations generated during WI. Some
details of the complete set of perturbation equations considered in WI
have been given in the Appendix~\ref{appB}. In the cold inflation
scenario, i.e., in the absence of dissipative effects and no radiation
bath during inflation, the primordial scalar curvature power spectrum
$\Delta_{{\cal R}}$ and the primordial tensor power spectrum
$\Delta_{T}$ are given by the standard expressions~\cite{Lyth:2009zz}
\begin{eqnarray} \label{PkCI}
\Delta_{{\cal R}} &=&  \left(\frac{ H^2}{2 \pi\dot{\phi}}\right)^2 ,
\\ \Delta_{T} &=& \frac{2 H^2}{\pi^2 M_{\rm Pl}^2},
\label{tensor}
\end{eqnarray}
respectively.  Because of dissipation and the presence of a radiation
bath, the primordial scalar power spectrum  given by Eq.~(\ref{PkCI})
changes, while the tensor spectrum Eq.~(\ref{tensor}) remains
unchanged.  The primordial power  spectrum for WI at horizon crossing
is given by \cite{Ramos:2013nsa,Benetti:2016jhf}
\begin{equation} 
\Delta_{{\cal R}}(k/k_*) =  \left(\frac{ H_{*}^2}{2
  \pi\dot{\phi}_*}\right)^2  {\cal F} (k/k_*),
  \label{Pk}
\end{equation}
where the function ${\cal F} (k/k_*)$ in Eq.~(\ref{Pk}) is 
(see Appendix~\ref{appB} for details)
\begin{equation}
{\cal F} (k/k_*) \equiv  \left(1+2n_* + \frac{2\sqrt{3}\pi
  Q_*}{\sqrt{3+4\pi Q_*}}{T_*\over H_*}\right) G(Q_*),
\label{calF}
\end{equation}
where $n_*$ denotes the thermal distribution of the inflaton field due
to the presence of the radiation bath and $G(Q_*)$ accounts for the
effect of the coupling of the inflaton and radiation
fluctuations~\cite{Graham:2009bf,BasteroGil:2011xd,Bastero-Gil:2014jsa}.  
The function $G(Q_*)$, in general, can
only be determined by numerically solving the set of perturbation
equations in WI.  The specific form for the function $G(Q_*)$ depends
mostly on the type of dissipation coefficient appearing in a WI model
and weakly on the inflaton potential chosen, at least for $Q \gg 1$.  
The tensor-to-scalar ratio $r$ and the spectral tilt $n_s$,  are defined 
in a standard way as 
\begin{equation}
r= \frac{\Delta_{T}}{\Delta_{\cal R}},
\label{eq:r}
\end{equation}
and
\begin{equation}
n_s -1 = \lim_{k\to k_*}   \frac{d \ln \Delta_{{\cal R}}(k/k_*) }{d
  \ln(k/k_*) }.
\label{eq:n}
\end{equation}
A subindex $*$ means that the quantities are evaluated at the Hubble
radius crossing of the pivot scale $k_*$  ($k_*=a_* H_*$).  The Planck
Collaboration~\cite{Akrami:2018odb} puts an upper bound on the
tensor-to-scalar ratio $r$ as $r<0.056$ (95$\%$ CL, Planck
TT,TE,EE+lowE+lensing+BK15, at the pivot scale $k_p = 0.002\,{\rm
  Mpc}^{-1}$), while the spectral tilt  is measured to be
$n_s=0.9658\pm 0.0040$  (95$\%$ CL, Planck
TT,TE,EE+lowE+lensing+BK15+BAO+running) at pivot scale $k_*=0.05\,{\rm
  Mpc}^{-1}$.   {}Furthermore, the normalization of the primordial
scalar curvature power spectrum, at the pivot scale $k_*$, is given by
$\ln\left(10^{10} \Delta_{{\cal R}} \right) \simeq 3.047$
(TT,TE,EE-lowE+lensing+BAO 68$\%$ limits), according to the Planck
Collaboration~\cite{Aghanim:2018eyx} and this is the value we will
assume in all our numerical simulations, in particular for finding the
normalization $V_0$ of the potential in Eq.~(\ref{pot}).

\section{Analytical determination of the field ranges}
\label{sec3}

As discussed in the Introduction, in this work, we are interested in
studying WI for the class of runaway potentials given by the
generalized exponential potentials of the form of Eq.~(\ref{pot}) with
$n>1$. As already pointed out in Ref.~\cite{Lima:2019yyv} (see also
Ref.~\cite{Das:2020lut}), for a simple functional form for the
dissipation coefficient in terms of the temperature $T$ and the inflaton 
field amplitude $\phi$, given by\footnote{
For some earlier studies also considering this functional form for the dissipation
coefficient in WI, see, e.g., Refs.~\cite{Zhang:2009ge,Visinelli:2016rhn}. }
\begin{equation}
\Upsilon(\phi,T)=C_\Upsilon\,T^p \phi^c M^{1-p-c},
\label{Upsilon}
\end{equation}
where $C_\Upsilon$ is a constant and $M$ some appropriate mass scale
associated with the microscopic model leading to Eq.~(\ref{Upsilon}) and using
the approximations in Eq.~(\ref{approx-eqns}), which are valid during
the WI dynamics, we find that the dissipation ratio $Q$ evolves with
the number of efolds $N$ like,
\begin{eqnarray}
\frac{d \ln Q}{dN} &=& - \frac{n \alpha \left(\frac{\phi}{M_{\rm
      Pl}}\right)^{n-2} } {4 - p + (4 + p) Q} \nonumber \\ &\times&
\left[ -2 p (n-1)-4c +(p-2) n \alpha \left(\frac{\phi}{M_{\rm
      Pl}}\right)^n \right].  \nonumber \\
\label{dQdNexpn}
\end{eqnarray}
Since the power $p$ in the temperature satisfies  $-4 < p <
4$ (see Refs.~\cite{Moss:2008yb,delCampo:2010by,BasteroGil:2012zr}),
thus, we find that only those cases of dissipation coefficient having
a power in the temperature with $p>2$ will lead to a dissipation ratio
decreasing  with the number of e-folds, which ensures that WI can
gracefully exit in these class of models~\cite{Das:2020lut}. 
Having a decreasing dissipation ratio $Q$ is also crucial in our derivation 
that follows below and which will allow 
us to work with very
steep potentials, yet keeping consistency with observations, in the large
dissipation regime of WI. 
For
instance, WI models with a dissipation coefficient $\Upsilon \propto
T^3$ fit this condition. 
A dissipation coefficient
with a cubic dependence in the temperature is also particularly well motivated by both
early and recent models of WI (see Appendix~\ref{appA} for some examples of particle
physics quantum field theory interaction schemes leading to this type
of dissipation coefficient)
and, therefore, it is quite
suitable for the study we have performed in the present work.
Thus, henceforth we will consider in all
of our analysis a dissipation coefficient  given simple by 
\begin{equation}
\Upsilon = C_\Upsilon \frac{T^3}{M^2}.
\label{cubic}
\end{equation}

\subsection{Background dynamics with the generalized exponential
  potential in WI}

\begin{center}
\begin{figure}
  \includegraphics[width=7.5cm]{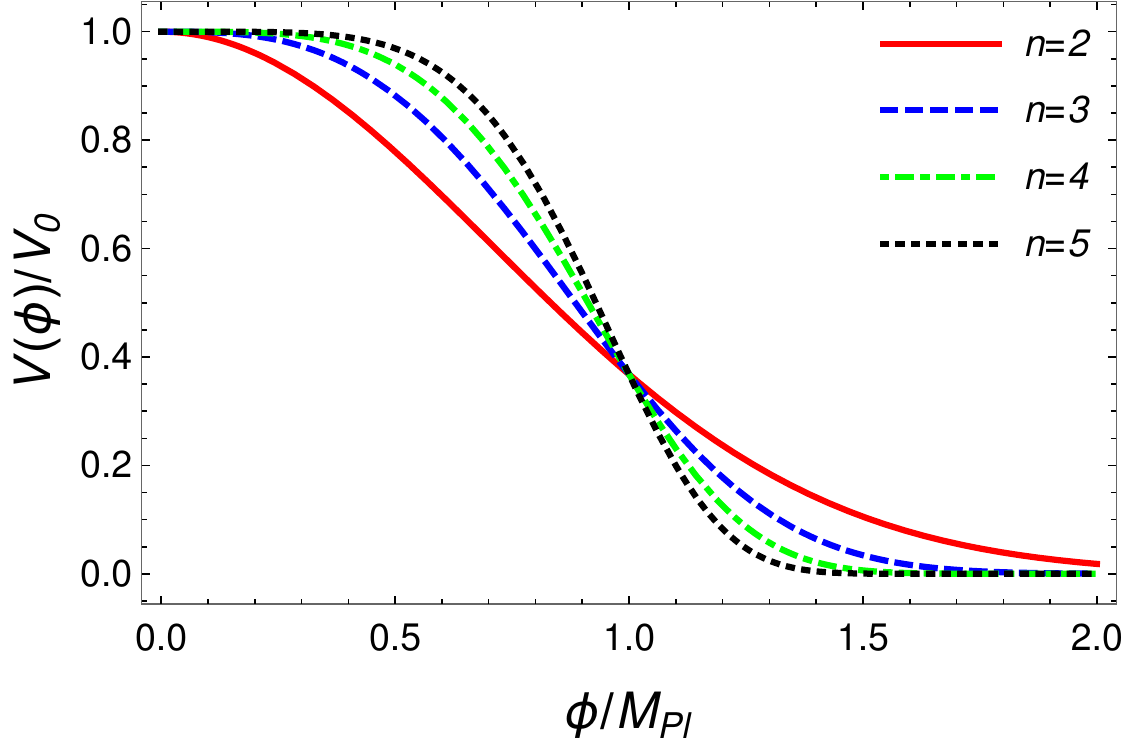}
  \caption
 {The form of the generalized exponential potential (\ref{pot}) for
   different values of $n$ and by choosing $\alpha=1$.  }
  \label{fig1}
\end{figure}
\end{center}

The dynamics of the model given by the potential in Eq.~(\ref{pot}) 
can be divided in two
regimes, depending on which region of the potential inflation starts
and ends.  The potential Eq.~(\ref{pot}) has a plateau region at
around $\phi=0$ and  an inflection point located at (for $n>1$)
\begin{equation}
\frac{\phi_{\rm inflection}}{M_{\rm Pl}} = \left(\frac{n-1}{n \alpha}
\right)^{1/n}. 
\end{equation}
In the region $0 < \phi < \phi_{\rm inflection}$, the dynamics is
similar to that of a hilltop inflation \cite{Boubekeur:2005zm}. WI happening in this region
favors the weak dissipative regime, $Q \ll 1$, as shown for hilltop
type of potentials in general~\cite{Benetti:2016jhf} (this was also
the regime studied in Ref.~\cite{Lima:2019yyv} for the generalized
exponential potential).  On the other hand, when WI happens entirely
in the runaway part of the potential, $\phi > \phi_{\rm inflection}$,
the strong dissipative regime of WI is favored. This is the part of
the potential we are interested to explore in this work, as
motivated by the swampland program. The form of the potential in
Eq.~(\ref{pot}), for different cases for the exponent $n$, is shown in
{}Fig.~\ref{fig1}. 

Let us now confirm that WI can indeed gracefully exit in the runaway
part of the potential.  WI ends when $\epsilon_V\approx 1+Q$, or
$\epsilon_{WI}\approx 1$. Therefore, to end inflation, $\epsilon_{WI}$
should be an increasing function of number of $e$-foldings yielding a
condition~\cite{Das:2020lut}
\begin{eqnarray}
\frac{d\ln\epsilon_V}{dN}>\frac{Q}{1+Q}\frac{d\ln Q}{dN}. 
\label{end-inf-cond}
\end{eqnarray}

{}For the runaway potential in Eq.~(\ref{pot}) the slow-roll
parameters become
\begin{eqnarray}
\!\!\!\!\!\!\!\!\!\epsilon_V&=&\frac{\alpha^2n^2}{2}\left(\frac{\phi}{M_{\rm
    Pl}}\right)^{2n-2},\\ 
\!\!\!\!\!\!\!\!\!\eta_V &=&\alpha^2n^2\left(\frac{\phi}{M_{\rm Pl}}\right)^{2n-2}
\!\!\!-\alpha
n(n-1)\left(\frac{\phi}{M_{\rm Pl}}\right)^{n-2}.
\end{eqnarray}
The slow-roll parameter $\epsilon_V$ evolves with the number of
$e$-foldings as
\begin{eqnarray}
\frac{d\ln\epsilon_V}{dN}=\frac{4\epsilon_V-2\eta_V}{1+Q}=\frac{\alpha
  n(n-1)}{1+Q}\left(\frac{\phi}{M_{\rm Pl}}\right)^{n-2}.
\label{eqdlnepsilon}
\end{eqnarray}
This shows that $\epsilon_V$ is a constant and does not evolve when
$n=1$, and it does evolve for $n>1$.  We are interested in the
part of the potential for which $\phi>\phi_{\rm inflection}$. We then 
get from Eq.~(\ref{eqdlnepsilon}) that
\begin{eqnarray}
\left(\frac{\phi}{M_{\rm
    Pl}}\right)^{n}=\frac{d\ln\epsilon_V}{dN}\frac{1+Q}{\alpha
  n(n-1)}\left(\frac{\phi}{M_{\rm Pl}}\right)^{2}.
\end{eqnarray}
The left hand side of the above equation has to be greater than
$(n-1)/n\alpha$ such that inflation takes place in the steep slope of
the potential. Thus, in this range we get 
\begin{eqnarray}
\frac{d\ln\epsilon_V}{dN}>\frac{(n-1)^2}{1+Q}\left(\frac{\phi}{M_{\rm
    Pl}}\right)^{-2}.
\end{eqnarray}
As the right hand side of the above inequality is always positive,
$\epsilon_V$ would always increase when inflation is taking place in
the steep region of the slope. Thus, inflation is assured to end
whenever $Q$ is constant or decreases with $e$-foldings. However,
inflation can also end when $Q$ increases with $N$ but with a slower
rate than the evolution of $\epsilon_V$ as shown in
Eq.~(\ref{end-inf-cond}).  This is, however, a more difficult
condition to achieve in general.

With the form of the dissipative coefficient given in
Eq.~(\ref{cubic}), the dissipation ratio $Q$ evolves as 
\begin{eqnarray}
\frac{d\ln Q}{dN}=\frac{10\epsilon_V-6\eta_V}{1+7Q}. 
\end{eqnarray}
In the strong dissipative regime ($Q\gg1$) the condition to end
inflation given in Eq.~(\ref{end-inf-cond}) then becomes 
\begin{eqnarray}
9\epsilon_V>4\eta_V,
\end{eqnarray}
which yields 
\begin{eqnarray}
\left(\frac{\phi}{M_{\rm Pl}}\right)^{n}>-\frac{8(n-1)}{n\alpha }.
\end{eqnarray}
As we are only interested in the region beyond the inflection point,
this condition will always be satisfied in the region of our
interest. Thus, inflation will always end in the steep potential
region beyond the inflection point. 

\subsection{The perturbations}

Now, let us consider the perturbations in this theory. The form of the
scalar power spectrum in WI has already been discussed in
Eq.~(\ref{Pk}) and Eq.~(\ref{calF}). In this case, the form for the
function $G(Q_*)$, valid for the generalized exponential potential
Eq.~(\ref{pot}), is found to be given by (see also discussion
concerning this point in the Appendix~\ref{appB}).
\begin{eqnarray}
G(Q_*) &=& \frac{ 1+6.12 Q_*^{2.73} }{ \left(1+6.96
  Q_*^{0.78}\right)^{0.72} } \nonumber \\ &+& \frac{0.01
  Q_*^{4.61}\left(1+ 4.82\times 10^{-6} Q_*^{3.12} \right)}
{\left(1+6.83 \times 10^{-13} Q_*^{4.12} \right)^2},
\label{GQ}
\end{eqnarray}
which is found to hold up to rather large values of $Q_*$ ($Q_* \sim 3\times
10^3$). This will be enough to cover the range of values of $Q$ to be
considered in the next section.

Note that the scalar power spectrum used in Ref.~\cite{Berghaus:2019whh} was of
the form (see, e.g., Eq.~(12) in that reference),
\begin{eqnarray}
\Delta_{\mathcal R}\approx
\frac{\sqrt{3}}{4\pi^{\frac32}}\frac{H^3T}{\dot\phi^2}
\left(\frac{Q}{Q_3}\right)^9Q^{\frac12},
\label{ps-old}
\end{eqnarray}
where $Q_3\approx 7.3$.  
It can be easily shown that in the strong dissipative regime a more accurate 
form of the power spectrum given in Eq.~(\ref{Pk}), along with the relations given in
Eq.~(\ref{calF}) and Eq.~(\ref{GQ}), does not differ much from the above equation. 
As the above equation is presented in a much simpler form for the strong dissipative regime,
we will derive the scalar
spectral index analytically using the form of the power spectrum
given in Eq.~(\ref{ps-old}) for simplicity (while numerically we will
use the full form of the function $G(Q)$ given in Eq.~(\ref{GQ}) to
calculate the spectral index). We see that 
\begin{eqnarray}
n_s-1=3\frac{d\ln H}{dN}-2\frac{d\ln\dot\phi}{dN}+\frac{d\ln
  T}{dN}+\frac{19}{2}\frac{d\ln Q}{dN}.\nonumber\\
\end{eqnarray}
To determine the quantities on the right hand side of the above
equations, we will use the approximated background equations given in
Eqs.~(\ref{approx-eqns}), along with the approximated forms of $Q$ and
$T$ valid for the form of the dissipation coefficient given by
Eq.~(\ref{cubic}) (see also Ref.~\cite{Berghaus:2019whh}),
\begin{eqnarray}
Q^7\approx \tilde C_Q\frac{V'^6}{V^5},\quad T^7\approx \tilde
C_T\frac{V'^2}{V^{1/2}},
\end{eqnarray}
which are valid in a strong dissipative regime. {}From these
equations, we see that 
\begin{eqnarray}
&&\frac{d\ln H}{dN}\sim-\frac{\epsilon_V}{Q}, \quad \frac{d\ln
    Q}{dN}\sim\frac{10\epsilon_V-6\eta_V}{7Q},\nonumber\\ &&\frac{d\ln
    T}{dN}\sim\frac{\epsilon_V-2\eta_V}{7Q}, \quad
  \frac{d\ln\dot\phi}{dN}\sim-\frac{3\epsilon_V+\eta_V}{7Q}.
\end{eqnarray}
Inserting all these in the expression of $n_s$, we
get~\cite{Berghaus:2019whh}
\begin{eqnarray}
n_s-1=\frac{3}{7Q}(27\epsilon_V-19\eta_V),
\end{eqnarray}
It is to note that, to obtain a red-tilt of the scalar spectrum one
requires a potential yielding $\eta_V>\epsilon_V$, as has also been
observed in Ref.~\cite{Berghaus:2019whh}. To be more precise, we need
$19\eta_V>27\epsilon_V$, which yields a condition 
\begin{eqnarray}
\left(\frac{\phi}{M_{\rm Pl}}\right)^n>\frac{38(n-1)}{11n\alpha}.
\end{eqnarray}
This is a stronger bound on the field range than the one required for
field ranges beyond the inflection point. Therefore, beyond the above
range, inflation ends as well as we get the desired spectral index
with appropriate choices of $\alpha$ for a given $n$. These findings
are explicitly checked in our numerical examples considered in the next 
section.

\section{Numerical analysis of the parameter ranges}
\label{numerical}

We have numerically evolved the full background equations given in
Eqs~(\ref{full-eqns}) for the cases $n=2$ to 5 and the findings are
furnished in the Table~\ref{tab1}. In principle, we can
tune appropriately both $Q_*$ and the constant $\alpha$ of the potential,
for a given value of $n$, such as to produce results consistent with the
observable quantities for either smaller or larger values of $Q_*$ than the
ones shown in Table~\ref{tab1}.  Our criteria for choosing the value
of $Q_*$ was that it would be large enough such that all the swampland
criteria could be met as also to have $n_s$ around the central value
from the Planck analysis. The tensor-to-scalar ratio $r$ is naturally very much
suppressed in the large $Q$ regime of WI, as already seen in other 
cases (see, e.g., Refs.~\cite{Bastero-Gil:2019gao, Kamali:2019xnt}).
The second and third
column containing the values of $n_s$ and $r$, respectively ensures
that the models with different values of $n$ (and the chosen values of
$\alpha$ accordingly) are in accordance with the observations. The
fifth column shows the amount of field traversed, $\Delta \phi$,
which turns out to be sub-Planckian for all the cases and it confirms that these
examples obey the Swampland distance conjecture. The tenth and the
eleventh column, which contain the values of the slow-roll parameters
$\epsilon_V$ and $\eta_V$, respectively, which are much larger than
unity, ensure that the Swampland de Sitter conjecture is
maintained. The last column, which quotes the values of $V_*$, i.e.,
the scale of inflation, confirms that the TCC is obeyed as
well. Hence, Table~\ref{tab1} confirms that the runaway potentials
with $n>1$ not only remain in tune with observations but also obey all
the three Swampland conjectures, making these models prime candidates
as consistent effective field theory models in String Landscapes.

\begin{widetext}
\begin{center}
\begin{table}[!htb]
\begin{tabular}{c|c|c|c|c|c|c|c|c|c|c|c}
\hline \hline Model & $n_s$ & $r$ & $N_*$ & $\Delta \phi/M_{\rm Pl}$ &
$C_\Upsilon$ & $M/M_{\rm Pl}$ & $T_{\rm end}$ (GeV) & $V_0$ (GeV)$^4$&
$\epsilon_{V_*}$ & $\eta_{V_*}$  & $V_*^{1/4}/M_{\rm Pl}$\\ \hline
$n=2$ &  &  &  &  &  &  &  & & & &\\ $\alpha= 9.6$& 0.9648 &
$4.89\times 10^{-29}$ & 48.0 & 0.98 & $6.04 \times 10^{-11}$ &
$4.0\times 10^{-13}$ & $1.52 \times 10^7$ & $3.07\times 10^{38}$ &
31.7 & 44.2 & $1.52\times 10^{-11}$\\ $Q_*=850.96$ &  &  &  &  &  &  &
& & & &\\ \hline $n=3$ &  &  &  &  &  &  &  & & & &\\ $\alpha= 2.5$&
0.9689 & $1.44\times 10^{-28}$ & 48.2 & 0.97 & $4.19 \times 10^{-11}$
& $4.0\times 10^{-13}$ & $2.49\times 10^{7}$ & $1.82\times 10^{39}$ &
25.9& 37.2  & $2.49\times 10^{-11}$\\ $Q_*=740.15$ &  &  &  &  &  &  &
& & & & \\ \hline $n=4$ &  &  &  &  &  &  &  & & & &  \\ $\alpha=0.45$
& 0.9655 & $1.83\times 10^{-28}$ & 48.1 & 0.97 & $3.86 \times
10^{-11}$ & $4.0\times 10^{-13}$ & $2.86\times 10^{7}$ & $3.59\times
10^{39}$ & 25.0& 36.5 & $2.85\times 10^{-11}$\\ $Q_*= 719.68$ &  &  &
&  &  &  &  & & & & \\ \hline $n=5$ &  &  &  &  &  &  &  & & & &
\\ $\alpha=0.06$ & 0.9645 & $2.37 \times 10^{-28}$ & 48.2 & 0.97 &
$3.53 \times 10^{-11}$ & $4.0\times 10^{-13}$ & $3.18\times 10^{7}$ &
$6.01\times 10^{39}$ & 24.1 & 35.6 & $3.18\times 10^{-11}$\\ $Q_*=
699.53$ &  &  &  &  &  &  &  & & & & \\ \hline \hline
\end{tabular}
\caption{Numerical estimation of respective parameters and the
  relevant cosmological quantities obtained from them.}
\label{tab1}
\end{table}
\end{center}
\end{widetext}

\begin{center}
\begin{figure}
  \includegraphics[width=7.5cm]{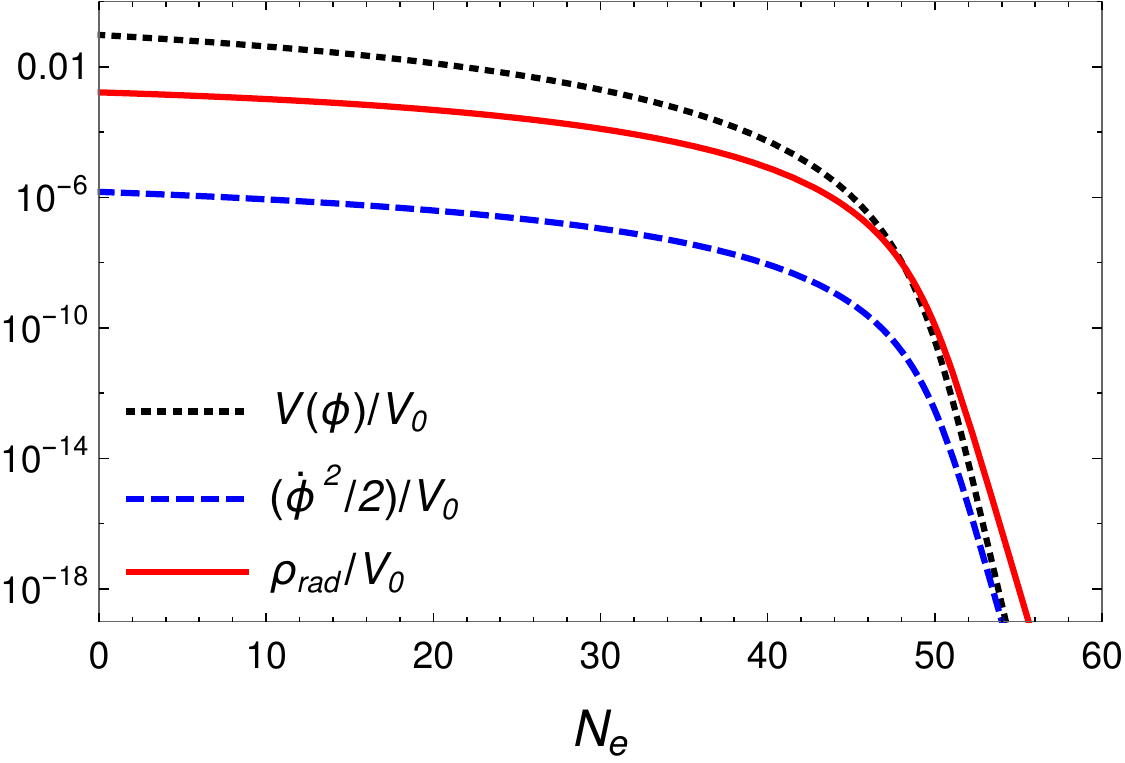}
  \caption
 {The evolution of the potential, kinetic and radiation energy
   densities for the model with $n=3$ given in Table~\ref{tab1}.  }
  \label{fig2}
\end{figure}
\end{center}

{}For completeness, we have also shown in {}Fig.~\ref{fig2} how the
potential, kinetic and radiation energy densities evolve with the
number of $e$-foldings for the $n=3$ case. The situation for the other
models with $n=2,\, 4$ and 5 are very much similar, so we do not show
them explicitly here. It can be seen that inflation ends when the
radiation energy density starts to dominate over the potential energy
density, while the kinetic energy density remains small all along.

As an additional point to be commented, and also related to the result
shown in {}Fig.~\ref{fig2}, is the number 
of efolds shown in the fourth column of Table~\ref{tab1}. As noticed from
{}Fig.~\ref{fig2}, WI exists gracefully at around 48
$e-$foldings, when the energy scale of inflation is as low as
$10^{-11}\, M_{\rm Pl}$. In WI we can precisely
determine the number of e-folds of inflation $N_*$  between the moment
the relevant  scales with wavenumber $k_*$ leaving the Hubble radius
and reentering around today. This is due to the fact that WI is not
followed by a reheating phase and, thus, there is no uncertainty
associated with the total number of $e$-folds which generally appears
due to the uncertainty in the number of $e$-foldings of the reheating
phase. By relating the comoving Hubble scale at $N_*$ when the mode
with comoving wavenumber $k_*$ crossed the horizon, $a_* H_*=k_*$,
with the one at present time, $a_0 H_0$, we have
that~\cite{Liddle:2003as}
\begin{equation}
\frac{k_*}{a_0 H_0} = \frac{a_*}{a_{\rm end}} \frac{a_{\rm
    end}}{a_{\rm reh}} \frac{a_{\rm reh}}{a_0}  \frac{H_*}{H_0},
\label{scales}
\end{equation}
where $a_*/a_{\rm end}= \exp(-N_*)$ and we can use the fact that at
the end of WI the universe smoothly transits to the radiation
dominated regime with no intermediary reheating phase, hence, $a_{\rm
  end}/a_{\rm reh} = 1$. This is particularly well satisfied for all
the models exemplified in Table~\ref{tab1}. {}Furthermore, as
also seen in {}Fig.~\ref{fig2}, inflation ends when $V(\phi)$
drops below $\rho_{R}$, while the kinetic energy density $\dot
\phi^2/2$ remains always subdominant, even after inflation.
Therefore, there is no kination period as is observed in general for
these type of runaway exponential potentials. {}Finally, we can relate
$a_{\rm end} \equiv a_{\rm reh}$ with $a_0$ in Eq.~(\ref{scales}) by
assuming that after WI there is no additional sources of entropy and
use then the entropy conservation result,
\begin{equation}
g_s(T_{\rm end}) T_{\rm  end}^3 a_{\rm end}^3 = \left( 2 T_{0}^3 +
\frac{21}{4} T_{\nu,0}^3 \right) a_0^3,
\end{equation}
where $T_0$ and $T_{\nu,0}=(4/11)^{1/3}T_0$ are, respectively, the
today's (CMB) photon and neutrino  temperatures and we have explicitly
used the respective number of degrees of freedom, while $g_s(T_{\rm
  end})$ is the effective number of degrees of freedom at the end of
WI.

Putting the above relations together, Eq.~(\ref{scales}), in WI,
becomes
\begin{equation}
\frac{k_*}{a_0 H_0} = e^{-N_*} \left[ \frac{43}{11 g_s(T_{\rm end})}
  \right]^{1/3} \frac{T_0}{T_{\rm end}} \frac{H_*}{H_0}.
\label{N*}
\end{equation}
Here, we take the convention that $a_0=1$. {}For the Hubble parameter
today,  we assume the Planck result, $H_0=67.66\, {\rm km}\, s^{-1}
{\rm Mpc}^{-1}$ [from the Planck Collaboration~\cite{Aghanim:2018eyx},
  TT,TE,EE-lowE+lensing+BAO 68$\%$ limits,  $H_0 = (67.66 \pm 0.42)\,
  {\rm km}\, s^{-1} {\rm Mpc}^{-1}$].  Likewise, for the CMB
temperature today we assume the value $T_0 = 2.725\, {\rm K}= 2.349
\times 10^{-13}\, {\rm GeV}$, while the for the pivot scale $k_*$ we
take the Planck value $k_* = 0.05/{\rm Mpc}$.  {}For $g_s(T_{\rm
  end})$ we assumed the standard model value, $g_s(T_{\rm end}) =
106.75$, for definiteness (the results are very weakly dependent on
$g_s$). Putting all these values in the above
equation, $N_*$  turns out to be of order 48, consistent with the
results seen in Table~\ref{tab1}. 

As a final remark about the observational predictions of the models
studied here,  concerns the level of non-Gaussianity that they produce.
{}For WI in the large dissipative regime, as is the case for all models
we have considered, it is predicted a non-Gaussianity coefficient of
the warm shape (see, e.g., Ref.~\cite{Bastero-Gil:2014raa} for details)
as large as $|f_{\rm NL}^{\rm warm}| \sim 5$ in the case of a dissipation coefficient
$\Upsilon \propto T^3$. This is still within the
range of the results obtained by the Planck 2018 team using the WI 
shape~\cite{Akrami:2019izv}, $f_{\rm NL}^{\rm warmS}=-48\pm 27$ 
(from SMICA+T+E, $68\%$ C.L.) and   $f_{\rm  NL}^{\rm warmS}=-39\pm 44$ 
(from SMICA+T, $68\%$ C.L.). But the expected result that we have for $f_{\rm NL}$, 
for all the cases here studied, is still in a magnitude that can be large 
enough to possibly be probed from, e.g., the future fourth
generation CMB observatories~\cite{Abazajian:2016yjj}, 
or from future large scale structure surveys.  
Both of these are expected to bring down the present upper bounds on 
non-Gaussianities. Thus, non-Gaussianity can be one of the indicators 
differentiating WI in the strong dissipative regime from the
weak dissipative one, through their distinct non-Gaussianity shapes
and predictions for $f_{\rm NL}$~\cite{Bastero-Gil:2014raa}.

\section{Conclusion}
\label{conclusions}

The proposed Swampland conjectures, namely the Swampland Distance
conjecture, the de Sitter conjecture and the Transplanckian Censorship
Conjecture, have severely constrained the construction of viable
inflation models in any String Landscape.  Thus, the swampland program
strongly restricts the class of possible effective field theory models
of inflation that are consistent with quantum gravity. It is thus
worthwhile to look for constructions where the inflationary dynamics
can be accommodated away from Swamplands. It has been pointed out
previously that the WI scenario befits the Swampland Conjectures,
especially the de Sitter one, much better than its counter part, the
cold inflation paradigm. The de Sitter conjecture is also better
suited with the runaway type of potentials (like, e.g.,  Eq.~(\ref{pot})) which
do not have any critical points. In particular, it has already been
demonstrated that the de Sitter conjecture also remains robust in
WI~\cite{Brandenberger:2020oav}.  Besides of this, it has been a
challenge, even for WI, to satisfy the TCC, which in its original
formulation~\cite{Bedroya:2019tba}, requires inflation to happen at
sufficiently small scales. The difficulty is associated with the
construction of WI models able to support strong enough dissipation
and at the same time to be consistent with the
observations~\cite{Bastero-Gil:2019gao}.  Even models motivated by WI,
like the ones studied in Ref.~\cite{Berera:2020iyn}, reflects well
this difficulty.  It is thus an
important task finding appropriate inflation models that are able to
satisfy all
the swampland criteria.  In the present paper, we have explicitly
considered the validity of runaway type of potentials when embedding
them in WI. 

It was previously shown that the runaway potential with exponent $n=1$
can be embedded in a Randall-Sundrum braneworld inflation successfully
where it can observe all the three Swampland
conjectures~\cite{Kamali:2019xnt}. However, when such a potential is
embedded in a standard general relativity context with WI, it yields
too much red tilt in the scalar power spectrum to be in accordance
with the observations~\cite{Goswami:2019ehb}.  In this work we have
examined the runaway potentials with exponents $n>1$ when
embedded in WI models characterized by a cubic in the temperature
dependence of the dissipation coefficient to show that (a) such models
gracefully exists inflation when inflation takes place in the runaway
part of the potential; (b) they can remain in tune with the current
observations by yielding the correct scalar spectral index, and (c)
they can also simultaneously satisfy all the three Swampland
conjectures as a consequence of supporting a strong enough dissipative
regime of WI. The combination of all these features make these type of
models viable inflation models when constructed in the WI picture and
which can be realized within String landscapes.

\section*{Acknowledgments}

The work of S.D. is supported by Department of Science
and Technology,  Government of India under the Grant Agreement number
IFA13-PH-77 (INSPIRE Faculty Award).  R.O.R. is partially supported by
research grants from Conselho Nacional de Desenvolvimento
Cient\'{\i}fico e Tecnol\'ogico (CNPq), Grant No. 302545/2017-4, and
Funda\c{c}\~ao Carlos Chagas Filho de Amparo \`a Pesquisa do Estado do
Rio de Janeiro (FAPERJ), Grant No. E-26/202.892/2017.
 

\appendix

\section{The dissipation coefficient}
\label{appA}

Dissipative effects are expected to be experienced by systems,  when
displaced from their state of equilibrium and interacting with  an
environment. We can consider the case of a background scalar field
$\phi$  initially displaced from its equilibrium state and
interacting with other fields $X$. Given an interaction Lagrangian
density like
\begin{equation}
{\cal L}_{\rm int}(\phi, X) = -f(\phi) g(X).
\label{VphiX}
\end{equation}
A proper study of the evolution of the background field can be
performed in the context of the in-in, or the closed-time path (CTP)
functional formalism~\cite{Calzetta:2008iqa}.  By integrating over the
$X$ field, a nonlocal effective equation of motion for $\phi$ can
be derived and the ensemble  averaged effective equation of motion for
$\phi$ can be generically expressed like~\cite{Calzetta:2008iqa}
\begin{widetext}
\begin{eqnarray}
\partial_\mu \frac{\partial {\cal L}_{\rm eff,r}[\phi]} {\partial
  (\partial_\mu \phi)} - \frac{\partial   {\cal L}_{\rm
    eff,r}[\phi]} {\partial \phi} - i \frac{\partial
  f(\phi)}{\partial \phi} \int d^4 x' \theta(t-t') \left[
  f(\phi(x')) - f(\phi(x)) \right] \langle [ g(X(x)),g(X(x')) ]
\rangle =  0,
\label{effeom}
\end{eqnarray}
\end{widetext}
where ${\cal L}_{\rm eff,r}[\phi]$ is the renormalized effective
Lagrangian density for $\phi$ and $\langle \cdots \rangle$ are
ensemble averages with respect to an equilibrium (quantum or thermal)
state. Equation~(\ref{effeom}) form the basis of the many earlier
works~\cite{Gleiser:1993ea,Berera:1998gx,Berera:2001gs,Berera:2004kc,Berera:2007qm}
(for a review, see also Ref.~\cite{Berera:2008ar}) that evolved to
warm inflation model realizations.  The non-local term in
Eq.~(\ref{effeom}) represents a transfer of energy from the $\phi$
field into radiation. The nonlocal term in  Eq.~(\ref{effeom}) can be
localized and expressed in a form of a proper dissipation term when
there is a separation of timescales between the system and
environment, e.g., given a response timescale $\tau$ related to the
plasma interactions  and $\phi$ is slowly varying on the response
timescale $\tau$, $\dot \phi/\phi \ll \tau^{-1}$, which is
typically referred to as the adiabatic approximation, then we can
write~\cite{Berera:2007qm}
\begin{widetext}
\begin{eqnarray}
-i \frac{\partial f(\phi)}{\partial \phi} \int d^4 x'
\theta(t-t') \left[ f(\phi(x')) - f(\phi(x)) \right] \langle [
  g(X(x)),g(X(x')) ] \rangle \approx \Upsilon  \dot\phi,
\label{eomdiss1}
\end{eqnarray}
\end{widetext}
where $\Upsilon$ is the dissipation coefficient defined as
\begin{equation}
\Upsilon=\int d^4 x' \Sigma_R (x,x') (t'-t) = -\lim_{\omega\to 0}
\frac{{\rm Im} \Sigma_R({\bf k}={\bf 0},\omega)}{\omega},
\label{UpsilonSigmaR}
\end{equation}
where $\Sigma_R(\omega)$ is the {}Fourier-transform of the retarded
correlation function,
\begin{equation}
\Sigma_R (x,x') = -i\left[ \frac{\partial f(\phi)}{\partial
    \phi} \right]^2 \theta(t-t') \langle [ g(X(x)),g(X(x')) ]
\rangle.
\label{Sigmacorr0}
\end{equation}

Many examples of dissipation coefficients were for example derived in
Ref.~\cite{BasteroGil:2010pb}.  As discussed in the Introduction, in
this work we are particularly interested in models leading to a
dissipation coefficient that scales with the cubic power in the
temperature, $\Upsilon \propto T^3$. Let us briefly review viable
interaction schemes leading to such a dissipation coefficient.

\subsection{Dissipation through a catalyst heavy field}

One of the first working field theory model for WI has
been constructed in the case of the inflaton field dissipating to
light radiation fields intermediate by a heavy catalyst field. The
implementation is based on a supersymmetric model with  chiral
superfields $\Phi$, $X$ and $Y_i$, $i=1,\ldots,N_Y$, described by the
superpotential~\cite{Berera:2008ar},
\begin{equation} \label{superpotential}
W={g\over2}\Phi X^2+{h_i\over2} XY_i^2+f(\Phi)~,
\end{equation}  
where a sum over the index $i$ is implicit.  The scalar component of
the superfield  $\Phi$ describes the inflaton field, with an
expectation value $\phi=\phi/\sqrt{2}$ and $f(\Phi)$ describes the
self-interactions in the inflaton sector.   The $X$ fields are assumed
to be heavy fields with respect to the radiation bath temperature
produced by the light fields $Y$, i.e., $m_X \gg T$ and $m_Y \ll
T$. Under these circumstances, the dissipation coefficient in the
inflaton's equation of motion can be shown to be given
by~\cite{BasteroGil:2010pb,BasteroGil:2012cm}
\begin{equation}
\Upsilon \simeq C_\Upsilon \frac{T^3}{\phi^2}, \;\;\; C_\Upsilon
\simeq \frac{\alpha_h}{4} N_X N_Y,
\label{susy}
\end{equation}
where $\alpha_h=h^2N_Y/(4 \pi)$, $N_X$ is the number of fields in the
$X$ heavy sector and it was assumed $h=h_i$ for all the light fields
for simplicity.  despite the dependence on the inflaton field in
Eq.~(\ref{susy}), the results obtained for this model do not differ
much from the ones we have obtained in Sec.~\ref{numerical}.  It can
likewise support strong dissipation and satisfy all the swampland
criteria and with an observationally  consistent $n_s$.  However, for
$Q \sim {\cal O}(700)$ as in Table~\ref{tab1}, we have instead that
$C_\Upsilon \sim 10^{14}$, which, from Eq.~(\ref{susy}) and assuming
$h \sim 1$, it implies the need of a huge number of heavy and/or light
fields, $N_X N_Y \sim 5 \times 10^{15}$. Such a large number might be
a technical challenge associated with this model, from both the
perturbativity and the unitarity point of view, associated with this
model for the present analysis (see, however,
Ref.~\cite{BasteroGil:2011mr} for a possible scenario where these
issues can be overcame and that uses brane constructions, or also
the proposal in Ref.~\cite{Matsuda:2012kc} where large field
multiplicities can be allowed due to a Kaluza-Klein tower in
extra-dimensional scenarios).

\subsection{Dissipation in the Minimal Warm Inflation model}

The Minimal Warm Inflation (MWI) model was proposed by the authors in
Ref.~\cite{Berghaus:2019whh}.  In the MWI model the inflaton field has
axion-like couplings to non-Abelian gauge fields which yields a viable
model of the thermal bath that can exist during WI. The inflaton field
is coupled to a Yang-Mills field $A_\mu^a$ in an axion-like form,
\begin{equation}
{\cal L}_{\rm int} = \frac{\alpha_g}{8 \pi} \frac{\phi}{f}
\tilde{F}^{a\,\mu\nu} F^a_{\mu \nu},
\label{axionL}
\end{equation}
where the dual gauge field strength $\tilde{F}^{a\,\mu \nu} = \frac{1}{2}\epsilon^{\mu \nu \alpha \beta}
F^a_{\alpha\beta}$, $F_{\mu\nu}^a = \partial_\mu A_\nu^a - \partial_\nu
A_\mu^a  + g C^{abc} A_\mu^b A_\nu^c$, with $g$ the Yang-Mills
coupling and $C^{abc}$ is the structure constant of the non-Abelian
group. In Eq.~(\ref{axionL}) one also has that $\alpha_g \equiv g^2/(4
\pi)$ and $f$ is an scale analogous to the axion decay constant.
The corresponding dissipative coefficient that the interaction
produces has been shown to be  related to the Chern-Simons diffusion
rate and given by~\cite{Moore:2010jd,Laine:2016hma} 
\begin{eqnarray}
\Upsilon= C_\Upsilon \frac{T^3}{f^2},\;\;\;  C_\Upsilon
=\kappa(\alpha_g, N_c, N_f) \alpha_g^5,
\label{upsilonaxion}
\end{eqnarray}
where $N_c$ is the dimension of the gauge group, $N_f$ is the
representation of the fermions if any, and $\kappa$ is a dimensional
quantity depending on $N_c$, $N_f$ and $\alpha_g$. 

One of the advantages of this model is that the shift symmetry
satisfied by the inflaton naturally protects it from large thermal
corrections that might  undermine the slow-roll conditions during
inflation. 

It is useful to estimate the scale $f$ appearing in
Eq.~(\ref{upsilonaxion})  from our numerical results given in
Table~\ref{tab1}. Using that $\kappa \sim 100$ and $g\sim 0.1$, then
from the values for $C_\Upsilon$ and $M$ given in Table~\ref{tab1}, we
find that for all models analyzed $f\sim 5 \times 10^4$ GeV. It is
clear that in the present context we cannot associate $\phi$ with the quantum
cromodynamics (QCD) axion, since $f$ is much below the astrophysical
lower  bound $f \gtrsim 10^9$ GeV set in for the QCD axion decay
constant~\cite{Sikivie:2020zpn}.

\subsection{Dissipation through derivative couplings with the inflaton field}

A third option to produce dissipation coefficients behaving like
$\Upsilon \propto T^3$ is motivated from the previous example. We
consider the case where the inflaton has a moduli-like (or
dilaton-like) derivative coupling with other radiation fields, which
can be for example other scalar fields and with an interaction
Lagrangian density given by
\begin{equation}
{\cal L}_{\rm int} = g^2 \frac{\phi}{2M} (\partial_\mu \chi)^2.
\label{moduliL}
\end{equation}
The scalar field $\chi$ is supposed to remain in thermal equilibrium
through either its self-coupling or to couplings to other radiation
fields (e.g., additional gauge or other light fields that could be added to the
model). It has been realized in Ref.~\cite{Bodeker:2006ij} that the
dissipation coefficient in this model can be precisely related to the bulk
viscosity calculation for a scalar field~\cite{Jeon:1994if}, leading
then to the result for the dissipation coefficient given by
\begin{equation}
\Upsilon \simeq 4.5 g^4 \frac{\ln^2(\xi \lambda)}{\lambda}
\frac{T^3}{M^2},
\label{upsilonmoduli}
\end{equation}
where $\xi$ is a numerical constant, $\xi= \exp[15 \zeta(3)/\pi^2]
\simeq 0.064736$, and $\lambda$ is the quartic self-coupling for the
$\chi$ field, $-\lambda \chi^4/4!$.  It is interesting to observe that
this connection of the dissipation coefficient in WI at high
temperature with a viscosity coefficient was already noticed
in Ref.~\cite{Berera:1998gx}.

An additional interaction that now makes the connection with the bulk
viscosity calculation in a pure gauge theory is by coupling the moduli
field now to pure Yang-Mills gauge fields through a coupling like
(note that this is different from the interaction term in
Eq.~(\ref{axionL})
\begin{equation}
{\cal L}_{\rm int} = -g^2 \frac{\phi}{2M} F^{a\,\mu\nu} F^a_{\mu \nu},
\label{moduliL2}
\end{equation}
which gives for the dissipation coefficient the
result~\cite{Laine:2010cq,Laine:2016hma}
\begin{equation}
\Upsilon \simeq \frac{(12 \pi \alpha_g)^2}{\ln(1/\alpha_g)}
\frac{T^3}{M^2},
\label{upsilonmoduli2}
\end{equation}
where, as before, $\alpha_g\equiv g^2/(4 \pi)$.

It remains, as possible future work, to see how the above moduli-like
interactions can be implemented in an explicit quantum field theory
model building construction for WI.

\section{Perturbations in warm inflation}
\label{appB}

We review here the first-order perturbation equations for WI,
consisting of the inflaton perturbations $\delta \phi$, the radiation
energy density perturbation $\delta \rho_r$ and the radiation momentum
perturbation $\Psi_r$.  The notation that we follow is the one given in
Refs.~\cite{Hwang:1991aj,Hwang:2001fb}.

The  perturbed FLRW metric  is given by
\begin{eqnarray}
ds^2 &=& -(1+2 \alpha) dt^2 - 2 a \partial_i \beta dx^i dt   \nonumber
\\ &+& a^2 [ \delta_{ij} (1 +2 \varphi) + 2 \partial_i \partial_j
  \gamma] dx^i dx^j, \label{metric}
\end{eqnarray}
where $\alpha$, $\beta$, $\gamma$ and $\varphi$ are the
spacetime-dependent  perturbed-order variables.  These metric
perturbation functions are related to the complete set of equations
(when {}Fourier transforming to
space-momentum)~\cite{Hwang:1991aj,Hwang:2001fb}
\begin{eqnarray}
&& \chi = a ( \beta + a \dot \gamma) \,, 
\label{chi} \\
&&\kappa= 3 (H \alpha - \dot \varphi) + \frac{k^2}{a^2} \chi \,, 
\label{kappa} \\
&&-\frac{k^2}{a^2} \varphi + H \kappa = - \frac{1}{2 M_{\rm Pl}^2}
\delta \rho\,,
\label{varphi}\\
&& \kappa -\frac{k^2}{a^2} \chi = - \frac{3}{2 M_{\rm Pl}^2} \Psi\,,
\label{kappachi}\\
&&\dot{\chi} + H \chi -\alpha -\varphi = 0\,,
\label{dotchi}\\
&&\dot{\kappa} + 2 H \kappa + \left(3 \dot{H} - \frac{k^2}{a^2}
\right) \alpha = \frac{1}{2 M_{\rm Pl}^2} \left( \delta \rho + 3\delta
p \right)\,,
\label{dotkappa}
\end{eqnarray}
where $\delta \rho$, $\delta p$ and $\Psi$ are, respectively,  the
total density, pressure and momentum perturbations. In our two-fluid
system (inflaton plus radiation), they are given in terms of the 
inflaton field and radiation perturbations, e.g.,
\begin{eqnarray}
&&\delta \rho = \delta \rho_\phi + \delta\rho_r \,,
\label{deltarho}\\
&& \delta p = \delta p_\phi + \delta p_r\,,
\label{deltap}\\
&& \Psi = \Psi_\phi + \Psi_r\,,
\label{Psi}
\end{eqnarray}
with $\delta \rho_\phi = \dot{\phi} \delta\dot{\phi} - \dot{\phi}^2
\alpha + V_{,\phi} \delta \phi$, $\delta p_\phi = \dot{\phi}
\delta\dot{\phi} - \dot{\phi}^2 \alpha - V_{,\phi} \delta \phi$,
$\delta p_r = \omega_r \delta \rho_r$ and $\Psi_\phi = - \dot{\phi}
\delta \phi$ (with ``dot'' always denoting   derivative with respect
to the cosmic time).

The evolution equations for the field and radiation perturbation
quantities follow from the conservation of the energy-momentum
tensor. The complete equations have been given in
Ref.~\cite{BasteroGil:2011xd}, which we indicate the interested 
reader for more details.
Working in momentum space, defining the {}Fourier transform with
respect to the comoving coordinates, the equation of motion for the
radiation and momentum fluctuations with comoving wavenumber  $k$ are
given by   
\begin{eqnarray}
&& \delta \dot{\rho_{r}}+4H\delta\rho_{r} =(1+\omega_r)\rho_{r}
  \left(\kappa-3H \alpha\right) \nonumber \\ &&  +
  \frac{k^{2}}{a^{2}}\Psi_{r}+\delta Q_{r}+Q_{r}\alpha \,, 
\label{energyalpha}
\\ && \dot{\Psi}_{r}+3H\Psi_{r}+\omega_r\delta \rho_{r}+\delta \Pi =
-(1+\omega_r)\rho_{r}\alpha+J_{r} ,  \nonumber \\
\label{momentumalpha} 
\end{eqnarray}
where 

\begin{eqnarray}
&&Q_r  = \Upsilon \dot \phi^2 , \label{Qr}\\  &&\delta Q_r = \delta
  \Upsilon \dot \phi^2 + 2 \Upsilon \dot \phi \delta \dot \phi - 2
  \alpha \Upsilon \dot \phi^2 \,, \label{deltaQr}\\  &&J_r = -
  \Upsilon \dot \phi \delta \phi .\label{Jr}
\end{eqnarray}

In addition to Eqs. (\ref{energyalpha}) and (\ref{momentumalpha}),
there is also the evolution equation for the field fluctuations
$\delta \phi$, which is described by a stochastic evolution determined
by a Langevin-like equation~\cite{Graham:2009bf}:
\begin{eqnarray}
&&\delta \ddot \phi + 3 H  \delta \dot \phi + \left(\frac{k^2}{a^2} +
  V_{,\phi\phi}\right) \delta \phi =  \xi_q+\xi_T  \nonumber - \delta
  \Upsilon \dot \phi \nonumber \\  && + \dot \phi ( \kappa + \dot
  \alpha) + (2 \ddot \phi + 3 H \dot \phi) \alpha  -\Upsilon ( \delta
  \dot \phi - \alpha \dot \phi) \label{field}\,,
\end{eqnarray}
where $\xi_{q,T}\equiv \xi_{q,T}({\bf k},t)$  are stochastic Gaussian
sources related to quantum and thermal fluctuations with appropriate
amplitudes (for details and for their complete definitions, 
see Ref.~\cite{Ramos:2013nsa}).
 
To complete the specification of the fluctuation equations, we need
$\delta \Upsilon$, the fluctuation of the dissipation coefficient.
{}For a general temperature $T$ and field $\phi$ dependent dissipative
coefficient, given by Eq.~(\ref{Upsilon}),
we have that
\begin{equation}
\delta \Upsilon =  \Upsilon \left[p \frac{\delta T}{T}+c \frac{\delta
    \phi}{\phi} \right]. \label{dupsilon} 
\end{equation}

Although dissipation implies departures from thermal equilibrium in
the radiation fluid, the system has to be close-to-equilibrium for the
calculation of the dissipative coefficient to hold, therefore we
assume $p_r \simeq \rho_r/3$ and, hence, $\omega_r=1/3$.  Then, with
$\rho_r \propto T^4$, we have that  $\delta T/T \simeq \delta
\rho_r/(4\rho_r)$ and $\delta Q_{r}$ in Eq. (\ref{deltaQr}) can be
expressed as
\begin{equation}
 \delta Q_{r}= 3 H Q \dot{\phi}^2 \left(\frac{p \delta \rho_{r}}{4
   \rho_r} +\frac{c \delta \phi}{\phi} \right) + 6 H Q \dot{\phi}
 \delta \dot{\phi} - 6 H Q \dot{\phi}^2 \alpha\, .
\label{deltaQalpha}
\end{equation}

{}From the above relations, the complete system of first-order
perturbation equations for WI become 
\begin{eqnarray}
\delta \ddot{\phi} &=& - 3H\left(1+Q\right)\delta \dot{\phi} -
\left(\frac{k^{2}}{a^{2}}+V_{,\phi\phi}+\frac{3c
  HQ\dot{\phi}}{\phi}\right)\delta \phi  \nonumber \\ &+& \xi_q+\xi_T
- \frac{p H}{\dot{\phi}}\delta \rho_{r}+
\dot{\phi}(\kappa+\dot{\alpha}) +
    [2\ddot{\phi}+3H(1+Q)\dot{\phi}]\alpha,  \nonumber \\
\label{deltaddotphi} 
\\ \delta \dot{\rho_{r}} &=& -H \left(4 - \frac{3pQ\dot{\phi}^2}{4
  \rho_r}  \right)\delta \rho_{r} + \frac{k^2}{a^2}\Psi_{r} +
6HQ\dot{\phi} \delta \dot{\phi}  \nonumber \\ &&+ \frac{3 c
  HQ\dot{\phi^{2}}}{\phi} \delta \phi + \frac{4  \rho_r}{3}  \kappa -
3 H\left( Q \dot{\phi}^2 + \frac{4\rho_r}{3}  \right) \alpha,
\nonumber \\
\label{deltadotrhor} 
\\ \dot{\Psi}_{r}  &=&- 3H \Psi_{r} - 3 H Q \dot{\phi} \delta \phi -
\frac{1}{3} \delta \rho_r   - 4 \rho_r \frac{\alpha}{3} . 
\label{dotpsir}
\end{eqnarray}
Equations~(\ref{deltaddotphi}), (\ref{deltadotrhor}) and
(\ref{dotpsir}), together with the metric perturbations
Eqs.~(\ref{chi}) - (\ref{dotkappa}),  form a complete set of equations
in a ``gauge-ready'' form. {}From this point on we can either choose to
work in terms of gauge-invariant quantities
\cite{Kodama:1985bj,Hwang:1991aj}, or equivalently just choose an
appropriate  gauge directly.  Even though any appropriate gauge can be
chosen, a convenient one showing good numerical stability when
numerically integrating the full set of differential equations is the
Newtonian slicing (or zero shear) gauge $\chi=0$.  In the $\chi=0$
gauge, the relevant metric equations become
\begin{eqnarray}
&&\kappa= \frac{3}{2 M_{\rm Pl}^2} ( \dot{\phi} \delta \phi - \Psi_r
  )\,,
\label{kappachi=0}\\
&& \alpha = -\varphi \,,
\label{alphachi=0}\\
&& \dot{\varphi} = - H \varphi -\frac{1}{3} \kappa\,.
\label{dotvarphichi=0}
\end{eqnarray}
{}Finally, the power spectrum is determined from the comoving
curvature perturbation  ${\cal R}$, defined as
\begin{equation}
\Delta_{\cal R}(k)= \frac{k^3}{2 \pi^2} \langle |{\cal R}|^2
\rangle\,, \label{PR}
\end{equation}
where ``$\langle \cdots \rangle$'' means average over different
realizations of the noise terms in Eq.~(\ref{deltaddotphi}) (see, for
instance
Refs.~\cite{Graham:2009bf,BasteroGil:2011xd,Bastero-Gil:2014jsa} for
details of the numerical procedure).  {}Finally, the comoving
curvature perturbation ${\cal R}$ is composed of contributions not
only from the metric perturbations and the inflaton momentum
perturbations, but also from the radiation momentum perturbations,
\begin{eqnarray}
&&{\cal R}= \sum_{i=\phi,r} \frac{\rho_i +
    \bar{p}_i}{\rho+\bar{p}}{\cal R}_i\;,
\label{R}\\
&&{\cal R}_i = - \varphi - \frac{H}{\rho_i + \bar{p}_i} \Psi_i\,,
\label{Ri}
\end{eqnarray}
with $\bar{p}=p_\phi + p_r$, $\bar{p}_\phi \equiv p_\phi$ and
$\bar{p}_r =p_r $.

Note that in the literature there are different forms for which the
resulting curvature  perturbations are presented. {}For instance, by
neglecting the explicit coupling between the inflaton and radiation
perturbations, e.g., by setting the temperature power of the
dissipation coefficient to zero, $p=0$, and dropping the metric
perturbations (which are first-order in the slow-roll coefficients),
Eq.~(\ref{deltaddotphi}) can be explicitly
solved~\cite{Ramos:2013nsa}, leading to the result, computed at Hubble
radius crossing $k=aH$,
\begin{eqnarray}
\Delta_{\cal R} &=& \frac{H^3T}{4\pi^2 \dot{\phi}^2}  \left[
  \frac{3Q}{2\sqrt{\pi}} 2^{2\alpha} \frac{\Gamma\left( \alpha
    \right)^2\Gamma\left( \nu -1\right)
    \Gamma\left(\alpha-\nu+3/2\right)}{\Gamma\left(\nu
    -\frac{1}{2}\right) \Gamma\left(\alpha+\nu-1/2\right)} \right.
  \nonumber \\ &+& \left.  \frac{H}{T}\coth\left(\frac{H}{2T} \right)
  \right],
\label{Pphi}
\end{eqnarray}
where $\nu = 3(1 + Q)/2$, $\alpha = \sqrt{ \nu^2 + 3\beta Q/(1+Q) - 3
  \eta_V}$, $\beta=M_{\rm Pl}^2 \Upsilon_{,\phi} V_{,\phi}/(\Upsilon
V)$ and $\Gamma(x)$ is the Gamma-function.  By dropping slow-roll
coefficients, $\alpha \approx \nu$ and Eq.~(\ref{Pphi}) can be very
well approximated by the result
\begin{eqnarray}
\!\!\!\!\!\!\!\!\!\!\!\!\!\Delta_{\cal R}&\simeq & \frac{H^4}{4\pi^2
  \dot{\phi}^2}\left[1+2n_{BE}(H/T) + \frac{2\sqrt{3}\pi
    Q}{\sqrt{3+4\pi Q}}{T\over H}\right],
\label{Pphi2}
\end{eqnarray}
where $n_{BE}$ is the Bose-Einstein distribution. In general
  we can also replace $n_{BE}$ by $n_*$, representing the statistical
  distribution state of the inflaton at Hubble radius crossing, which
  might not be necessarily that of thermal equilibrium.  The
form given by Eq.~(\ref{Pphi2}) is typically the result used in most
of the recent literature in WI.  When including the coupling between the
inflaton and radiation perturbations shown in  Eqs.~(\ref{deltaddotphi}),
(\ref{deltadotrhor}) and (\ref{dotpsir}), these equations can only be
solved numerically. The result is a correction to
e.g. Eq.~(\ref{Pphi2}) which can be expressed in the form of a
function  $G(Q)$ of the dissipation coefficient and determined by a
proper fitting of the  numerical result for the curvature
perturbation. In particular, for the cubic in the temperature
dissipation coefficient studied in this work, we obtain
Eq.~(\ref{GQ}). Note that there are varied ways in how the
  perturbation equations are solved, which lead to differences on how
  this function  $G(Q)$ is presented in the literature. {}For
  instance, in the Ref.~\cite{Graham:2009bf}, where this effect of the
  coupling between inflaton and radiation perturbations in WI was
  first studied, an approximation to $G(Q)$ was given by neglecting
  both metric perturbations and other terms proportional to slow-roll
  coefficients in the perturbation equations and only the leading
  order dependence on $G(Q)$ through a simplified fitting was
  presented. Simpler fittings were also presented in
  Ref.~\cite{BasteroGil:2011xd}.


\end{document}